# Capturing Broadband Light in a Compact Bound State in the Continuum

Zeki Hayran[1], Francesco Monticone[1,*]

[1] *School of Electrical and Computer Engineering, Cornell University, Ithaca, New York 14853, USA*

*\* Corresponding author: francesco.monticone@cornell.edu*

Trapping and storing light for arbitrary time lengths in open cavities is a major goal of nanophotonics, with potential applications ranging from energy harvesting to optical information processing. Unfortunately, however, the resonance lifetime of conventional open resonators remains finite even in the limit of vanishing material absorption, as a result of radiation loss. In this context, bound states in the continuum (BiCs) have provided a unique way to achieve unbounded resonance lifetimes despite the presence of compatible radiation channels. However, physical constraints such as reciprocity, linearity, and delay-bandwidth limits prevent the possibility to externally excite such ideal bound states and make them interact with broadband sources. Here, we overcome these limitations and theoretically demonstrate that subwavelength open resonators undergoing a suitable temporal modulation can efficiently capture a broadband incident wave into a nonradiating eigenmode of the structure, leading to the first example of a BiC that is accessible to broadband light. Our findings unveil the dynamic capabilities of bound states in the continuum and extend their reach and potential impact for different applications.

## 1. INTRODUCTION

In an open system, ideally localized waves or bound states (e.g., electrons in a potential well or optical modes in a waveguide) are typically separated, with respect to frequency/energy, from the continuous spectrum of extended radiation modes (e.g., outgoing plane waves or spherical waves). If a discrete bound state exists within the continuum, it will typically couple with symmetry-compatible radiation modes, becoming a so-called leaky mode or quasi-normal mode [1],[2]. The first counterexample to this behavior was discovered in 1929 by von Neuman and Wigner, who showed that, in certain engineered potentials, perfectly bound electronic states can co-exist with the radiation continuum, namely, they remain ideally confined despite having an energy larger than the potential barrier [3]. In recent years, these perfectly bound (i.e., non-radiating) states, later denoted as "bound states in the continuum" (BiCs) [4], have stimulated a great deal of research interest in photonics [5],[6],[7],[8],[9],[10],[11] for applications spanning from ultra-selective filters [12] and sensors [13],[14],[15], to lasers [16],[17]. Owing to their vanishing radiation loss and ultra-narrow spectral linewidth, such BiCs are especially promising in areas where high frequency selectivity and long interaction times are required [18]. However, due to their decoupling from the radiation continuum, ideal BiCs are extremely difficult to externally excite and control. This can be understood from simple reciprocity arguments: if an eigenmode of an open structure cannot decay via coupling with external radiation, then external incident fields cannot excite such a radiationless eigenmode [19]. A recent proposal to overcome these reciprocity restrictions is to use nonlinear materials with intensity-dependent refractive index, so that the structure is automatically "tuned" to the BiC condition as the internal field amplitude increases [20]. However, the use of nonlinearities implies a minimum threshold value for the incident field amplitude and the pulse duration, which might limit the applicability of this technique. Other studies have employed atomic nonlinearities to excite a BiC externally [21], [22]; however, the experimental realization of such schemes may be challenging due to the required precise control of the atom-cavity coupling. We should also stress that, while it is not possible to externally excite an ideal BiC, there is actually no limit, at least in theory, to how close the nonradiating condition can be approached. Quasi-BiCs can indeed be realized by detuning certain properties of the scatterer or the excitation field from the exact BiC condition, which allows exciting these long-lived resonances with an external incident field [23]. However, the resonance lifetime will inevitably degrade due to the additional radiation loss, thereby showing the same trade-offs of more conventional open cavities.

Indeed, another limitation quasi-BiCs have in common with conventional resonances is related to their interaction with broadband incident fields: the resonance bandwidth (range of frequencies that can interact with the quasi-BIC) cannot be widened without proportionally reducing the resonance lifetime (hence increasing the level of loss). This follows from a general resonance argument: since in an open optical resonator energy continuously leaks out while incident light is coupled into the resonator, the time length over which the field can be trapped (resonance lifetime) is inversely proportional to the rate at which the energy exits the resonator, which is in turn proportional to the linewidth of the resonance. This relation, which can be generalized to a much broader class of systems [24],[25],[26],[27] is commonly known as the delay-bandwidth limit (or time-bandwidth limit) and is the principal impediment for realizing permanent storage (i.e., infinite delay) for fields with non-zero bandwidth (any signal carrying genuine information must have non-zero bandwidth). In other words, it is impossible to force arbitrarily broadband fields into a long-lived resonance. In addition, while Lorentz reciprocity is the basic principle that forbids the external excitation of an ideal BIC [19], as discussed above, the delay-bandwidth limit applies to both reciprocal and nonreciprocal scenarios, as shown in [24], as long as the system is linear and time-invariant.

Given these general considerations, it is therefore relevant to wonder whether it is at all possible to realize an open cavity supporting a bound state in the continuum that is accessible to broadband light. If physically feasible, this would represent the largest possible violation of the delay-bandwidth limit since the product between signal bandwidth and interaction time would diverge. In this Article, inspired by recent works on perfect absorption in time-modulated cavities (e.g., [28],[29]), we show that these effects can indeed be achieved in suitably engineered time-varying scattering systems, and we theoretically demonstrate that a broadband signal can be forced into a nonradiating eigenmode.

## 2. RESULTS

### a. Theory

To keep our discussion as general as possible, we employ temporal coupled mode theory (CMT) [28],[31] to model the temporal dynamics of a generic, bounded (i.e., having finite size in all dimensions) open optical resonator that interacts with an impinging broadband pulse (see Fig. 1(a)). In this case, the CMT equations for a single-mode resonator coupled with an external optical port can be written as

$$\frac{dA(t)}{dt} = \left[i\omega_0(t) - \gamma_{ext}(t)\right]A(t) + \kappa(t)\psi_+(t) , \qquad (1)$$

$$\psi_-(t) = -\psi_+(t) + \kappa(t)A(t) , \qquad (2)$$

where $A(t)$ is the mode amplitude of the resonator, while $\psi_+(t)$ and $\psi_-(t)$ are the incoming and outgoing wave amplitudes at the external port, respectively. $\kappa(t)$ is the coupling coefficient between the resonator and the port, and $\omega_0(t) = \omega_m(t) + i\gamma_m(t)$ is the complex resonance frequency. Note that we have assumed that the system's parameters, the resonance frequency and the coupling coefficient, may vary in time. We have also added the term $\gamma_{ext}(t)$ to account for the possible growth/decay rate of the resonance mode due to a possible energy exchange between the mode and the external source of the modulation. We also note that previous CMT formulations for resonators with time-varying parameters [32],[33],[34],[35] employed a somewhat simplified form by assuming that a temporal perturbation in any of the system's parameters will not induce a change in the other parameters (for instance assuming that a time-modulation in $\omega_m$ will not affect $\gamma_m$ and $\kappa$ [35]). This is indeed a good approximation if the structural dimensions of the resonator are large compared to the free-space wavelength at the resonant frequency (as in the case of a conventional ring resonator [36]) and if the modulation region is limited to a specific, isolated region of the device. However, in this work we focus on compact open resonators that are potentially subwavelength in size and for which a generic parameter perturbation may alter the resonator dynamics dramatically. Therefore, we investigate the resonator interaction with the incident wave without any limiting assumption on any of the time-varying parameters.

Using time-reversal symmetry and energy conservation [28],[31], the coupling coefficient for the static (time-invariant) case can be shown to be related to the decay rate due to radiation leakage, which is equal to the inverse of the imaginary part of the complex resonance frequency, $\gamma_m$, if material absorption is zero (i.e., radiation is the only available decay channel). Thus, in the lossless case we obtain

$$\kappa = \sqrt{2\gamma_m}. \qquad (3)$$

Eq. (3) is typically assumed valid also for time-varying resonant structures, as done for example in [32],[33],[34],[35],[37]. In this context, however, it is important to note that in a time-varying optical system, conservation of energy within the system is broken since time-translation symmetry (time invariance) is locally broken [38] (the term $\gamma_{ext}(t)$ in Eq. (1) accounts for this possible external energy exchange). Even if the frequency transitions induced by a modulation are symmetric [32],[35] and energy is therefore ultimately conserved (the total decay and growth rates due to the modulation cancel each other out), the time-varying $\gamma_{ext}(t)$ may still affect the instantaneous parameters of the system during the modulation. Therefore, particular care must be taken when dealing with dynamic systems of this type, where seemingly exotic phenomena, such as negative extinction for certain system parameters [37], can occur due to the breaking of energy conservation. Nonetheless, one way to justify the validity of Eq. (3) for time-varying systems is to consider the energy exchange between the resonator and the external modulator as an additional optical port. As originally shown in Ref. [28], the coupling coefficient between the incident wave at a certain port and the resonator is derived based on the decay rate of the resonance mode at that port with the other ports disconnected. The introduction of the decay rate associated with the additional ports is then regarded as a small perturbation and is assumed to have a negligible effect on the coupling coefficient. Following this argument, we will consider Eq. (3) to be valid also for temporally modulated resonant systems, as usually done, and will further assume that the decay/growth rate of the resonator mode are dominated by the port-induced decay rate, namely, $\gamma_m(t) \gg \gamma_{ext}(t)$ in Eq. (1). This is because, although energy is locally not conserved, the total number of photons is still conserved throughout an optical modulation; hence, provided that the modulation does not alter the frequencies of the photons considerably, the mode amplitude decay due to radiation loss can be assumed to dominate the decay/growth due to the Doppler-like frequency shifts of the photons.

Following these considerations, we use Eqs. (1), (2) and (3) to study the conditions to achieve critical coupling (i.e., reflectionless light capturing) for broadband signals. Intuitively, critical coupling, i.e., $\psi_-(t) = 0$, is obtained when the direct reflection from the cavity and the "leakage" from the resonance mode destructively interfere in the input channel. This is impossible to achieve in a static single-port cavity with no internal absorption and with constant-amplitude incident fields (more on this point below), as all the input energy must be reflected backward to satisfy energy conservation in such a lossless scenario. As we will see in the following, the situation is very different for time-varying cavities and, most importantly, the critical coupling condition may be continuously maintained while the radiation leakage is reduced to zero, leading to the desired BiC behavior.

The dynamic equation for critical coupling, that is, for $\psi_-(t) = 0$, can be written as

$$\frac{d\psi_+(t)}{dt} = \left[ i\omega_m(t) + \gamma_m(t) + \frac{1}{2\gamma_m(t)} \frac{d\gamma_m(t)}{dt} \right] \psi_+(t) . \tag{4}$$

Eq. (4) can be solved from two different viewpoints: (i) Solving for the appropriate resonance frequency modulation $\omega_0(t)$ that guarantees critical coupling for a given incoming wave $\psi_+$; or (ii) solving for the appropriate incident wave $\psi_+$ that would undergo critical coupling for a given modulation $\omega_0(t)$. In the former case, Eq. (4) can be arranged as

$$\frac{d\gamma_m(t)}{dt} = 2\gamma_m(t) \left[ i[\omega_m(t) - \omega_i] + \gamma_m(t) + \frac{2(t-t_i)}{\Delta t^2} \right], \tag{5}$$

where $\psi_+$ is assumed to be a broadband Gaussian pulse given by $\psi_+(t) = \exp\left(i\omega_i t - (t-t_i)^2/\Delta t^2\right)$, where $\omega_i$ is the center frequency, $t_i$ is the time position of the pulse, and $\Delta t$ is related to its temporal width. Eq. (5) is particularly challenging to solve explicitly due to its nonlinear character and the fact that $\omega_m$ and $\gamma_m$ are generally related. Nevertheless, Eq. (5) may be solved for specific cases in which $\omega_m$ and $\gamma_m$ can be expressed in relatively simple analytical forms. For the second case, which requires solving for the appropriate incident wave, Eq. (4) becomes a linear differential equation for which an exact solution can be obtained as

$$\psi_+(t) = C\sqrt{\gamma_m(t)} e^{N(t)} , \quad N(t) = \int_{-\infty}^{t} \left( i\omega_m(t') + \gamma_m(t') \right) dt' , \tag{6}$$

where $C$ is a constant to be determined by a specified boundary condition. It can be seen from Eq. (6) that in the case of a static BiC with $\gamma_m = 0$, the critical coupling condition implies $\psi_+ = 0$, verifying the well-established fact that an ideal BiC cannot be externally excited. On the other hand, in the case of a static resonance mode with $\gamma_m \neq 0$, the solution reads $\psi_+(t) = m\sqrt{\gamma_m} \exp\left((i\omega_m + \gamma_m)t\right)$, which indicates that an exponentially growing incident wave is required to suppress the reflections from the resonator. This implies that critical coupling can be achieved even in the lossless case if the incident field is not constant in amplitude but, instead, grows exponentially in time, which corresponds to the recently proposed concept of virtual critical coupling [39],[40]. One way to intuitively understand this effect is that, in order to destructively interfere with the wave "leaking" from the exponentially growing resonator mode, the reflected wave and, therefore, the driving incident wave are required to have a similarly growing amplitude (clearly, however, the resonator starts reflecting as soon as the incident wave stops growing exponentially).

In the time-varying case, the solution of Eq. (6) with dynamic parameters is more delicate, since $\omega_m$ and $\gamma_m$ are usually strongly nonlinear functions of the modulation parameters. Hence, the required pulse profile, $\psi_+(t)$, to realize reflectionless light trapping is challenging to calculate explicitly. However, a heuristic inspection of Eq. (6) shows that the $\exp(N(t))$ factor is an exponentially growing function with a time-varying growth rate. Specifically, for the case of interest, the growth rate will decrease over time and will eventually converge to zero, since zero radiation loss, i.e., $\gamma_m \to 0$, is required to achieve a BiC. The exponential function will therefore saturate over time to a certain level. Multiplying such a function by the $\sqrt{\gamma_m(t)}$ coefficient in Eq. (6), which is a monotonically decreasing function of time, will then produce a Gaussian-like pulse shape for the required incident field $\psi_+$. Thus, as illustrated in Fig. 1(a), if a broadband approximately-Gaussian pulse impinges on the time-varying open cavity, modulated in such a way that the radiation leakage is suitably reduced over time, then the pulse will be captured with minimal reflection, since a Gaussian-like pulse guarantees critical coupling according to Eq. (6). Once the incident pulse is completely trapped inside the open resonant cavity, the decay rate for the mode of interest is made to converge to zero, thereby converting the resonance into a BiC and leading to permanent storage of the incident pulse. We stress that, as for any true BiC, the structure is effectively closed *only* from the point of view of the specific mode of interest, which is converted into a non-radiating mode with $\gamma_m = 0$ [19]. However, the structure remains "electromagnetically open," as non-zero internal fields can still be induced by a generic incident field if the resonator supports additional leaky modes (which is usually the case), as we will see in the following.

Figures 1(b-d) show an alternative picture of the process described above, with temporal modulations translated into frequency transitions [41],[42],[43],[44],[45],[46],[47],[48],[49], [50]. A broadband pulse composed of a continuum of states at different frequencies can be 'squeezed' into a single state at a certain discrete frequency (in this case a BiC) through a proper temporal modulation (see Fig. 1(b)). Physically, in the case of an extended cavity (having one or more infinite structural dimensions) supporting traveling-wave modes with a certain wavevector, this process can be interpreted as the result of direct photonic transitions, i.e., transitions with conserved wavevector in the band diagram of the structure [38]. Such frequency transitions are illustrated in Figs 1(c) and 1(d) for an extended cavity and a compact bounded resonator, respectively, supporting a BiC at a discrete value of wavevector and/or frequency (marked with white circles).

### b.  Design and Results

To numerically verify the proposed dynamic BiC excitation, we chose an open cavity formed by a resonant scattering object with finite size in all dimensions. The structure is "open" in the sense that it is not surrounded by perfectly reflecting mirrors that directly forbid outgoing waves. Realizing BiCs in a bounded open cavity of this type was thought to be difficult, if not impossible, since its quantum mechanical analogue corresponds to a compact finite potential, which cannot support an electronic bound state in the continuum (i.e., a bound state with positive energy) [8]. From a classical electrodynamics perspective, a compact, bounded, three-dimensional BiC would need to have zero electric and magnetic fields everywhere outside a spherical surface enclosing the structure, in order to ensure no outgoing power flow from the resonator (spherical harmonics are always radiative) [19]. As the electromagnetic boundary conditions and the analyticity of the fields cannot usually be satisfied in such a case without the fields being zero also inside the structure [6,9,10], a compact optical BiC of this type cannot be realized under normal circumstances. Nonetheless, it was demonstrated that this issue can be overcome using materials with vanishing permittivity $\varepsilon$ and/or permeability $\mu$ [9],[10],[20],[51], or certain nonlocal materials [52], which allow satisfying the boundary conditions with vanishing external fields and non-zero internal fields. Within this context, it was shown that a subwavelength-sized dielectric-metal core-shell resonator can be suitably designed to trap light indefinitely, at least in the limit of no material absorption, by suppressing the radiation leakage of the mode of interest when the permittivity of the shell material becomes zero [9],[10]. In other words, the internal polarization current of the mode of interest becomes a non-radiating source distribution at the BiC condition [19],[51]. Most importantly, the structure remains electromagnetically open even if the condition $\varepsilon = 0$ is ideally satisfied, namely, the scatterer is penetrable by generic incident fields [9],[10],[20],[51]. Such a structure is especially well suited to verify our theoretical predictions, since its subwavelength dimensions enable us to tune the resonator parameters over a wide range by perturbing a relatively small region of the resonator, namely, the shell. Another advantage is that fast modulation of the structure parameters can be achieved through optical pumps, which can externally alter the plasma frequency $\omega_p$ of the metallic material (or degenerately doped semiconductor) composing the plasmonic shell of the scatterer [53]. In addition, it was recently shown that tunability through optical beams becomes more efficient if the pump is around the zero-permittivity frequency of the material, as a consequence of the slow group velocity in this regime [54].Several recent works demonstrated sub-500 fs material excitation and relaxation times, which enabled ultrafast variations in $\omega_p$ [53],[55],[56] and adiabatic frequency shifting (redshift or blueshift) through varying the probe-pump delay [54],[57]. While an experimental demonstration goes beyond the scope of the present work, these recent results suggest that our proposal may be experimentally realized using analogous setups in similar frequency regimes (particularly around optical communication wavelengths) through suitably varying the probe-pump delay and the intensity and time duration of the pump pulse. The core-shell resonator under study can be designed to support an ideal BiC, in the limit of vanishing material loss, by overlapping the plasma frequency of the shell, at which $\varepsilon = 0$, with the resonance frequency of the cavity formed by the dielectric core surrounded by epsilon-near-zero walls, which act as perfect magnetic conductors for the transverse-magnetic (TM$^r$) dipolar mode of interest, as extensively discussed in [9],[10]. In other words, the mode can be converted from a leaky resonance into a BiC, and vice versa, by appropriately tuning $\omega_p$. Then, in order to investigate the tuning behavior of the resonator, we have determined the complex resonance frequency of the scatterer, for the considered eigenmode, based on exact analytical calculations [58] (Mie theory) as a function of $\omega_p$ and for various shell thicknesses. From these results, reported in Fig. 2, one may immediately note the markedly asymmetric behavior around the BiC frequency, $\omega_c$; in particular, the imaginary part of the complex resonance frequency, $\gamma_m$, is typically smaller for $\omega_p > \omega_c$. This can be understood as a consequence of the fact that the real part of the resonance frequency, $\omega_m$, is smaller than $\omega_p$ in this range, which implies that the shell exhibits negative permittivity at the eigenfrequency of the mode. Since a medium with negative permittivity is highly reflective due its purely imaginary wave impedance, the coupling between the resonance mode and free space will decrease, therefore decreasing $\gamma_m$, consistent with Eq. (3). Radiation leakage, however, cannot be reduced to identically zero unless the permittivity goes to zero (BiC) or to negative infinity (closed cavity), or if the thickness diverges.

Another relevant observation that can be made from Fig. 2 is that, while the shell thickness does not affect the real part of the eigenfrequency considerably, it strongly influences the imaginary part. This is expected because a thinner shell increases the overlap between the evanescent tails of the leaky mode and the free-space modes (outgoing spherical waves), which in turn will increase the radiation loss (thereby increasing $\gamma_m$). Note that exactly at the BiC frequency, $\omega_p = \omega_c$, the "penetration depth" of the resonance mode inside the shell becomes zero (the zero-permittivity shell expels any magnetic field, for the transverse-magnetic mode of interest [9,10]) (see Fig. 2(b)). Thus, the complex eigenfrequency becomes independent of the shell thickness, which can therefore be made, in principle, arbitrarily thin. Building upon these observations, in the following we will consider plasma-frequency modulations in the range $\omega_p \leq \omega_c$, applied to a scatterer with a relatively thin plasmonic shell ($d = 0.085\ \lambda_c$), which allows us to achieve large tunability with small perturbations.

To verify our theoretical predictions in the previous section, we first approach the problem from a time-reversed perspective: instead of exciting the modulated resonator with the appropriate incident wave $\psi_+(t)$, we time-reverse the modulation and allow $\psi_+(t)$ to emerge as an outgoing spherical wave from an initially excited resonator, and we study the associated dynamics. Fig. 3(a) shows a conceptual sketch of a core-shell spherical scatterer being modulated by an external control pump. Several examples of $\omega_p$ modulations, shown in Fig. 3(b), have been investigated with the goal to verify the accuracy of the CMT analysis for different modulation ranges and speeds. In all cases, the excited structure is gradually detuned from the BiC condition, leading to an increase in radiation leakage until the internal fields of the open cavity are reduced to zero. The modulations have been chosen to be continuous, with their time-derivatives being also continuous, to allow for a smooth and physical transition. Figs. 3(c,d) show the analytically calculated (using Eq. (6)) and numerically simulated temporal profiles of the internal fields, while Figs. 3(e,f) show the spectra of the emerging wave (which are approximately Gaussian as expected from our discussion in the previous section). The numerical results were obtained through finite-difference time-domain (FDTD) simulations with a 3D grid [59]. Comparing these results, we see that the CMT accurately models and predicts the cavity dynamics and the emerging wave spectrum.

The decay rate of the resonance mode is determined by the values of $\gamma_m$ during the plasma-frequency modulation. Taylor-expanding $\gamma_m(\omega_p)$ around $\omega_p = \omega_c$, and ignoring non-dominant terms, gives (for $d = 0.085\ \lambda_c$): $\gamma_m \approx \omega_c^{-1}(\omega_p(t) - \omega_c)^2$. The considered $\omega_p$ modulations are in the form: $\omega_p(t) = \omega_c + \frac{2q}{p^3}t^3 - \frac{3q}{p^2}t^2$, where $q$ is the modulation range (in units of $\omega_c$) and $p$ is the modulation time length. Substituting this expression into the Taylor expansion of $\gamma_m$ and taking the time integral yields

$$\int \gamma_m dt = \left(\frac{9q^2}{5p^4}t^5 - \frac{2q^2}{p^5}t^6 + \frac{4q^2}{7p^6}t^7\right)\omega_c^{-1} = \mathrm{Re}[N(t)]\ .$$

Then, according to Eq. (6), this implies that faster modulations (i.e., smaller $p$) and larger modulation ranges $q$ lead to a larger decay rate, $\mathrm{Re}[N(t)]$ (i.e., the resonance mode decays faster). As an illustrative example, comparing modulations numbered 1 (mod-1, blue) and 2 (mod-2, red) in Fig. 3, we note that the smaller $p$ value of mod-2 indeed translates into a faster decay compared to mod-1 (Figs. 3(c,d)). Moreover, comparing modulations numbered 2 and 4 (mod-4, green) reveals that the larger $q$ value of mod-4 implies a larger decay rate (Figs. 3(c,d)). Another important observation that can be made from Fig. 3 is that a five-fold decrease in the modulation range translates into approximately a two-fold decrease in the bandwidth (by comparing modulations 1 and 3 in Figs. 3(e,f)). This scaling behavior, which can be understood from Eq. (6), suggests that, although the bandwidth of the radiated wave is proportional to the modulation range, a significant reduction of this range (easier to implement in practice) does not prevent the ability of the time-varying structure to interact with relatively broadband signals.

We now return to our original scenario: dynamic excitation of a BiC by an external broadband wave pulse (the time-reversed counterpart of the scenario studied in the previous paragraphs). We note that, in general, particular or unusual spatial profiles for the incident wave are not necessary for our purposes, as the mode of interest can be excited as long as the incident wave contains the appropriate spherical harmonic (a simple plane wave, which contains all spherical harmonics, would be sufficient) [19]; however, the coupling efficiency will certainly depend on the specific type of excitation. We first consider the case of an incoming spherical wave [58],[60] having the same type of polarization and order (angular momentum) as the BiC that is being excited (in our case, order $n = 1$ and TM$^r$ polarization). The electric and magnetic field components of the incident wave at a particular frequency $\omega$ can be written as (for $r > R + d$) [58],[60],

$$\mathbf{E}_+ = -\frac{E_0}{\varepsilon_0}\nabla \times \mathbf{F} + \frac{iE_0}{\omega\mu_0\varepsilon_0}\nabla \times \nabla \times \mathbf{A}, \quad \mathbf{H}_+ = -\frac{E_0}{\mu_0}\nabla \times \mathbf{A} - \frac{iE_0}{\omega\mu_0\varepsilon_0}\nabla \times \nabla \times \mathbf{F}$$

$$\mathbf{A} = \hat{r}\frac{\cos\varphi}{\omega}\frac{3i}{2}k_0 r h_1^{(1)}(k_0 r) P_1^1(\cos\theta), \quad \mathbf{F} = \hat{r}\frac{\sin\varphi}{\eta_0\omega}\frac{3i}{2}k_0 r h_1^{(1)}(k_0 r) P_1^1(\cos\theta), \qquad (7)$$

where $E_0$ is the complex amplitude, $P_1^1(.)$ and $h_1^{(1)}(.)$ are the associated Legendre polynomials of first degree and the spherical Hankel function of the first kind, respectively, both of order 1. $k_0 = \omega/c$ ($c$ being the speed of light) and $\eta_0$ are, respectively, the wave number and the wave impedance of free space. The spatial profile of the excitation wave described by Eq. (7), weighted at each frequency by the corresponding amplitude of the Fourier transform of the temporal profile given by Eq. (6) (approximately Gaussian), will ensure that, at all times and all positions, the temporal dynamics of the incident wave and the temporal modulation will lead to critical coupling. In other words, the radiation leakage from the excited resonance mode will destructively interfere with the direct spherical reflection at the scatterer boundaries, thereby suppressing any outgoing wave and achieving critical coupling into the open cavity. Following this insight, FDTD numerical experiments of the broadband excitation of a BiC have been carried out and the results are shown in Figs. 4(a-d). An inspection of the time-domain response of the field inside the scatterer (Fig. 4(a)) shows that the energy gradually builds up until the end of the modulation, after which the internal field amplitude is virtually constant in time, verifying the excitation of a BiC with vanishing radiation leakage (see also the inset in Fig. 4(a) for the spectrum of the BiC field). We also note that the temporal field profile in Fig. 4(a) is indeed equal to the time-reversed version of the field profiles in Figs. 3(c,d), as expected. Figs. 4(b-d) provide a detailed look into the spatial field distributions at different time instants, verifying that no scattering occurs during the excitation (hence fulfilling the critical coupling condition). As further demonstrated in the corresponding time-domain animations in Supplemental Material†, the incident pulse is indeed fully captured by the open cavity with no outgoing wave being excited. Thus, rather strikingly, from the viewpoint of the broadband incident field the scatterer acts essentially as a perfect absorber (a matched load impedance), even though material absorption is assumed to be zero. This is similar to the behavior demonstrated in [28],[29] for monochromatic signals, but realized here for broadband pulses interacting with an open cavity (a scatterer in free space) supporting a bound state in the continuum, i.e., a radiationless eigenmode. Note also that, in this case, the product between the bandwidth of the incident signal and its interaction time with the resonator is in principle unbounded since we were able to force a broadband pulse into an infinitely long-lived resonance by compressing its bandwidth. To be more specific, while the cavity linewidth is still inversely proportional to the resonance lifetime at the end of the modulation, as they both depend on the imaginary part of the eigenfrequency, $\gamma_m$, the signal bandwidth that can be captured by the time-varying open resonator is independent of the initial or final value of $\gamma_m$, in stark contrast with static cavities [28]. We stress again that this is possible only because the system is not time-invariant. The structure is also nonreciprocal, but nonreciprocity alone has no bearing on the possibility of overcoming delay-bandwidth restrictions [24].

In practical applications, it might be challenging to generate an incident spherical-wave pulse with the required characteristics. Therefore, it is relevant to explore whether a dynamic BiC can be excited with a standard plane-wave pulse. In this regard, Figs. 4(e-h) show the FDTD results for the same time-varying scatterer as before under broadband plane-wave excitation. A plane wave consists of (i.e., it can be represented as) an infinite sum of spherical harmonics, not just the spherical wave corresponding to the electric dipolar BiC considered here. Thus, a plane wave illumination may excite additional (higher-order) modes of the scatterer, corresponding to an induced current distribution with different multipolar contributions (the number of non-negligible multipolar terms can be determined approximately as $k_0(R+d) \simeq 3$ [58]). These additional modes have different eigenfrequencies with respect to the dipolar mode in Fig. 2 and will be affected differently by the temporal modulation. In particular, these modes are not transformed into BiCs, but remain damped by radiation (with a decay rate perturbed by the temporal modulation). Hence, different from the case of a single spherical-wave excitation, for plane-wave illumination there is usually a non-negligible re-radiation of energy from these leaky modes, corresponding to non-zero scattering, as clearly seen in the time-snapshot of the field distribution in Fig. 4(g). The excitation of additional modes may also produce higher internal fields in the scatterer, as more energy is stored in different resonators; however, the resonances sustaining these higher fields are short-lived and decay rapidly. Nevertheless, the results in Fig. 4(e) show that, after an initial peak in intensity during the external excitation, the internal field amplitude converges to a value that is constant in time, verifying that a broadband plane-wave pulse can indeed excite an infinitely long-lived BiC with vanishing radiation leakage.

To gain further insight into the spatio-temporal field dynamics for the considered open cavity under plane-wave excitation, we have calculated time-resolved spatial field-amplitude distributions along the optical axis of the excitation (*x*-axis) (Fig. 5(a)). For comparison, we have additionally provided in Fig. 5(b) the time-resolved field profile for a static time-invariant open cavity (the same core-shell scatterer with no modulation), supporting a BiC at the same frequency $\omega_c$. Fig. 5(a) clearly demonstrates the broadband excitation of an infinitely long-lived resonance. While some re-radiation/scattering can be seen during the excitation period (forming a standing wave to the left of the scatterer in Fig. 5(a)), it is evident that, after the incident pulse and temporal modulation end, the field is permanently trapped inside the open resonator with virtually no decay through radiation. The results for the static resonator in Fig. 5(b), on the other hand, shows that the incident pulse cannot excite long-lived (or self-sustained) modes within the scatterer, and the internal fields decay almost immediately upon excitation. These results verify that the ideal BiC supported by this static structure cannot be externally accessed and excited due to reciprocity restrictions, consistent with the discussion in [19]. Moreover, the field profile in Fig. 5(b) confirms that the structure remains electromagnetically open at the BiC condition, as generic fields can penetrate the scatterer at this frequency (the spectrum of these internal fields is non-zero at the BiC frequency, as shown in the inset).

Lastly, we briefly explored the possibility to excite BiCs in non-spherical geometries under broadband illumination. In practical applications, such as energy harvesting or optical information processing, a BiC supported by a structure with arbitrary shape might indeed serve useful purposes. We have therefore shown in Figs. 6(a-f) an illustrative example in which an octahedron-shaped core-shell scatterer (see Fig. 6(a)) is externally illuminated by a broadband plane-wave pulse. Due to the difficulties in studying such a resonator analytically, we relied on numerical FDTD simulations to obtain the complex resonance frequencies of the mode of interest, as a function of the plasma frequency, and then used Eq. (6) to determine the incident wave and temporal modulation that guarantee reflectionless light trapping. The internal fields have been recorded in time and are reported in Fig. 6(b), clearly showing that a BiC can be excited in this non-canonical geometry, with a temporal dynamics similar to the spherical case in Fig. 4(e). The time-resolved field profile in Fig. 6(d) further confirms that, also in this case, the mode amplitude of the BiC is nearly constant after the modulation ends, which indicates ideal trapping of light in a non-spherical open resonator. We, therefore, conclude that the insight and design considerations provided by our theoretical analysis are valid for arbitrary geometries, allowing the broadband excitation of BiCs in structurally complex platforms.

## 3. DISCUSSION

In conclusion, we have theoretically shown the conditions required to capture broadband light pulses in time-varying open resonators supporting nonradiating eigenmodes. The theory was confirmed with FDTD full-wave simulations, demonstrating for the first time a BiC that can be externally excited with broadband light. This leads to an unbounded delay-bandwidth product, as a broadband signal is forced into an infinitely long-lived resonance, a possibility enabled by the time-varying nature of the scattering system. We have also shown that our theory can be conveniently extended to open resonators with non-canonical geometries, which may be more appealing for certain practical applications. We also note that, even though the wave interaction with the dynamic resonator involves a nonlinear process (frequency conversion), the impinging pulse can be scaled arbitrarily, as predicted by Eq. (6). Therefore, no requirement exists in this configuration for the incident-wave amplitude to surpass a certain threshold, in contrast with the case of BiCs in nonlinear intensity-dependent scatterers [20].

While the considered examples have focused on plasmonic-based BiCs, which would unavoidably suffer from the presence of material absorption around the plasma frequency of the metallic shell, we would like to stress that the ideas put forward in this article are general, and may also be applied to extended BiCs supported by lossless dielectric structures [8]. Furthermore, the proposed approach may also be used for the broadband excitation of quasi-BiCs with long but finite lifetimes (in this case $\gamma_m$ needs to converge to a non-zero value in Eq. (6)), which may be more easily implemented in practical scenarios. Indeed, as mentioned above, the signal bandwidth that can be captured by the time-varying scatterer is independent of the value of $\gamma_m$ at the beginning or end of the modulation.

In addition, while in the previous section we have suggested the possibility of using external optical pumps to dynamically modulate the plasmonic shell of the resonator, other modulation mechanisms may be employed to achieve the same goal [61] depending on the application. For instance, a possible alternative is to use tunable liquid crystals as the core material and employ all-optical [62], thermal [63] or electro-optic [64] tuning to modulate the resonance frequency of the resonator. Depending on the required modulation speed, these different mechanisms may be suitable for applications in different contexts and at different operational frequencies.

In summary, our results extend the reach of the concept of BiC and may help harness its full potential by lifting the restrictions of reciprocity and the delay-bandwidth limit. Given their generality, the ideas presented in this article may also be translated to different realms of wave physics, including acoustics, quantum wave mechanics, and elastodynamics.

**ACKNOWLEDGMENTS**

The authors acknowledge support from the National Science Foundation (NSF) with Grant No. 1741694, and the Air Force Office of Scientific Research with Grant No. FA9550-19-1-0043. Z.H. also acknowledges support through the Fulbright Foreign Student Program of the U.S. Department of State.

**REFERENCES**

1. F. Monticone, A. Alù, Leaky-wave theory, techniques, and applications: From microwaves to visible frequencies. *Proc. IEEE* **103**(5), 793-821 (2015), doi:10.1109/JPROC.2015.2399419.
2. P. Lalanne, W. Yan, K. Vynck, C. Sauvan, J. P. Hugonin, Light Interaction with Photonic and Plasmonic Resonances. *Laser Photonics Rev.* **12**(5), 1700113 (2018), doi:10.1002/lpor.201700113.
3. J. von Neumann, E. P. Wigner, On some peculiar discrete eigenvalues. *Phys. Zeitschrift* **30**, 467 (1929).
4. F. H. Stillinger, D. R. Herrick, Bound states in the continuum. *Phys. Rev. A* **11**(2), 446 (1975), doi:10.1103/PhysRevA.11.446.
5. D. C. Marinica, A. G. Borisov, S. V. Shabanov, Bound states in the continuum in photonics. *Phys. Rev. Lett*. **100**(18), 183902 (2008), doi:10.1103/PhysRevLett.100.183902.
6. E. N. Bulgakov, A. F. Sadreev, Bound states in the continuum in photonic waveguides inspired by defects. *Phys. Rev. B - Condens. Matter Mater. Phys.* **78**(7), 075105 (2008), doi:10.1103/PhysRevB.78.075105.
7. C. W. Hsu, B. Zhen, J. Lee, S. L. Chua, S. G. Johnson, J. D. Joannopoulos, M. Soljačić, Observation of trapped light within the radiation continuum. *Nature* **499**(7457), 188-191 (2013), doi:10.1038/nature12289.
8. C. W. Hsu, B. Zhen, A. D. Stone, J. D. Joannopoulos, M. Soljacic, Bound states in the continuum. *Nat. Rev. Mater.* **1**(9), 1-13 (2016), doi:10.1038/natrevmats.2016.48.
9. M. G. Silveirinha, Trapping light in open plasmonic nanostructures. *Phys. Rev. A - At. Mol. Opt. Phys.* **89**(2), 023813 (2014), doi:10.1103/PhysRevA.89.023813.
10. F. Monticone, A. Alù, Embedded photonic eigenvalues in 3D nanostructures. *Phys. Rev. Lett.* **112**(21), 213903 (2014), doi:10.1103/PhysRevLett.112.213903.
11. H. M. Doeleman, F. Monticone, W. Den Hollander, A. Alù, A. F. Koenderink, Experimental observation of a polarization vortex at an optical bound state in the continuum. *Nat. Photonics* **12**(7), 397-401 (2018), doi:10.1038/s41566-018-0177-5.
12. J. M. Foley, S. M. Young, J. D. Phillips, Symmetry-protected mode coupling near normal incidence for narrow-band transmission filtering in a dielectric grating. *Phys. Rev. B - Condens. Matter Mater. Phys.* **89**(16), 165111 (2014), doi:10.1103/PhysRevB.89.165111.
13. Y. Liu, W. Zhou, Y. Sun, Optical refractive index sensing based on high-Q bound states in the continuum in free-space coupled photonic crystal slabs. *Sensors* **17**(8), 1861 (2017), doi:10.3390/s17081861.
14. S. Romano, A. Lamberti, M. Masullo, E. Penzo, S. Cabrini, I. Rendina, V. Mocella, Optical biosensors based on photonic crystals supporting bound states in the continuum. *Materials* **11**(4), 526 (2018), doi:10.3390/ma11040526.
15. S. Romano, G. Zito, S. Torino, G. Calafiore, E. Penzo, G. Coppola, S. Cabrini, I. Rendina, V. Mocella, Label-free sensing of ultralow-weight molecules with all-dielectric metasurfaces supporting bound states in the continuum. *Photonics Res.* **6**(7), 726-733 (2018), doi:10.1364/prj.6.000726.
16. A. Kodigala, T. Lepetit, Q. Gu, B. Bahari, Y. Fainman, B. Kanté, Lasing action from photonic bound states in continuum. *Nature* **541**(7636), 196 (2017), doi:10.1038/nature20799.
17. S. T. Ha, Y. H. Fu, N. K. Emani, Z. Pan, R. M. Bakker, R. Paniagua-Domínguez, A. I. Kuznetsov, Directional lasing in resonant semiconductor nanoantenna arrays. *Nat. Nanotechnol.* **13**(11), 1042-1047 (2018), doi:10.1038/s41565-018-0245-5.
18. K. Koshelev, G. Favraud, A. Bogdanov, Y. Kivshar, A. Fratalocchi, Nonradiating photonics with resonant dielectric nanostructures. *Nanophotonics* **8**(5), 725-745 (2019), doi:10.1515/nanoph-2019-0024.

19. F. Monticone, D. Sounas, A. Krasnok, A. Alù, Can a Nonradiating Mode Be Externally Excited? Nonscattering States versus Embedded Eigenstates. *ACS Photonics* **6**(12), 3108-3114 (2019), doi:10.1021/acsphotonics.9b01104.
20. S. Lannebère, M. G. Silveirinha, Optical meta-atom for localization of light with quantized energy. *Nat. Commun.* **6**(1), 1-7 (2015), doi:10.1038/ncomms9766.
21. M. Cotrufo, A. Alù, Excitation of single-photon embedded eigenstates in coupled cavity–atom systems. *Optica* **6**(6), 799-804 (2019), doi:10.1364/optica.6.000799.
22. G. Calajó, Y. L. L. Fang, H. U. Baranger, F. Ciccarello, Exciting a Bound State in the Continuum through Multiphoton Scattering Plus Delayed Quantum Feedback. *Phys. Rev. Lett.* **122**(7), 073601 (2019), doi:10.1103/PhysRevLett.122.073601.
23. K. Fan, I. V. Shadrivov, W. J. Padilla, Dynamic bound states in the continuum. *Optica* **6**(2), 169-173 (2019), doi:10.1364/optica.6.000169.
24. S. A. Mann, D. L. Sounas, A. Alù, Nonreciprocal cavities and the time–bandwidth limit. *Optica* **6**(1), 104-110 (2019), doi:10.1364/optica.6.000104.
25. D. A. B. Miller, Fundamental limit to linear one-dimensional slow light structures. *Phys. Rev. Lett.* **99**(20), 203903 (2007), doi:10.1103/PhysRevLett.99.203903.
26. R. S. Tucker, P. C. Ku, C. J. Chang-Hasnain, Slow-light optical buffers: Capabilities and fundamental limitations. *J. Light. Technol.* **23**(12), 4046-4066 (2005), doi:10.1109/JLT.2005.853125.
27. J. B. Khurgin, "Bandwidth Limitation in Slow Light Schemes," in *Slow Light: Science and Applications*, J. Khurgin and R. S. Tucker, Eds. (Taylor & Francis Group, 2008), chap. 15.
28. D. L. Sounas, Virtual perfect absorption through modulation of the radiative decay rate. *Phys. Rev. B* **101**(10), 104303 (2020), doi:10.1103/PhysRevB.101.104303.
29. M. S. Mirmoosa, G. A. Ptitcyn, V. S. Asadchy, S. A. Tretyakov, Time-Varying Reactive Elements for Extreme Accumulation of Electromagnetic Energy. *Phys. Rev. Appl.* **11**(1), 014024 (2019), doi:10.1103/PhysRevApplied.11.014024.
30. H. Haus, "Coupling of Modes: Resonators and Couplers" in *Waves and Fields in Optoelectronics* (Prentice-Hall, 1984), chap. 7.
31. S. Fan, W. Suh, J. D. Joannopoulos, Temporal coupled-mode theory for the Fano resonance in optical resonators. *J. Opt. Soc. Am. A* **20**(3), 569-572 (2003), doi:10.1364/josaa.20.000569.
32. M. Minkov, Y. Shi, S. Fan, Exact solution to the steady-state dynamics of a periodically modulated resonator. *APL Photonics* **2**(7), 076101 (2017), doi:10.1063/1.4985381.
33. M. R. Shcherbakov, P. Shafirin, G. Shvets, Overcoming the efficiency-bandwidth tradeoff for optical harmonics generation using nonlinear time-variant resonators. *Phys. Rev. A* **100**(6), 063847 (2019), doi:10.1103/PhysRevA.100.063847.
34. M. R. Shcherbakov, K. Werner, Z. Fan, N. Talisa, E. Chowdhury, G. Shvets, Photon acceleration and tunable broadband harmonics generation in nonlinear time-dependent metasurfaces. *Nat. Commun.* **10**(1), 1-9 (2019), doi:10.1038/s41467-019-09313-8.
35. A. Dutt, M. Minkov, Q. Lin, L. Yuan, D. A. B. Miller, S. Fan, Experimental Demonstration of Dynamical Input Isolation in Nonadiabatically Modulated Photonic Cavities. *ACS Photonics* **6**(1), 162-169 (2019), doi:10.1021/acsphotonics.8b01310.
36. W. R. McKinnon, D. X. Xu, C. Storey, E. Post, A. Densmore, A. Delâge, P. Waldron, J. H. Schmid, S. Janz, Extracting coupling and loss coefficients from a ring resonator. *Opt. Express* **17**(21), 18971-18982 (2009), doi:10.1364/oe.17.018971.
37. M. R. Shcherbakov, R. Lemasters, Z. Fan, J. Song, T. Lian, H. Harutyunyan, G. Shvets, Time-variant metasurfaces enable tunable spectral bands of negative extinction. *Optica* **6**(11), 1441-1442 (2019), doi:10.1364/optica.6.001441.
38. C. Caloz, Z. L. Deck-Leger, Spacetime Metamaterials-Part II: Theory and Applications. *IEEE Trans. Antennas Propag.* **68**(3) 1583-1598 (2020), doi:10.1109/TAP.2019.2944216.
39. Y. Radi, A. Krasnok, A. Alu, Virtual Critical Coupling. *ACS Photonics* (2020), doi:10.1021/acsphotonics.0c00165.
40. S. Longhi, Coherent virtual absorption for discretized light. *Opt. Lett.* **43**(9), 2122-2125 (2018), doi:10.1364/ol.43.002122.

41. D. L. Sounas, A. Alù, Non-reciprocal photonics based on time modulation. *Nat. Photonics* **11**(12), 774-783 (2017), doi:10.1038/s41566-017-0051-x.
42. E. Ippen, S. Fan, J. Joannopoulos, Interband transitions in photonic crystals. *Phys. Rev. B - Condens. Matter Mater. Phys.* **59**(3), 1551 (1999), doi:10.1103/PhysRevB.59.1551.
43. K. Fang, Z. Yu, S. Fan, Photonic Aharonov-Bohm effect based on dynamic modulation. *Phys. Rev. Lett.* **108**(15), 153901 (2012), doi:10.1103/PhysRevLett.108.153901.
44. Z. Yu, S. Fan, Complete optical isolation created by indirect interband photonic transitions. *Nat. Photonics* **3**(2), 91 (2009), doi:10.1038/nphoton.2008.273.
45. A. Khorshidahmad, A. G. Kirk, Wavelength conversion by interband transition in a double heterostructure photonic crystal cavity. *Optics Letters* **34**(19), 3035-3037 (2009).
46. M. F. Yanik, S. Fan, Stopping Light All Optically. *Phys. Rev. Lett.* **92**(8), 083901 (2004), doi:10.1103/PhysRevLett.92.083901.
47. Q. Xu, P. Dong, M. Lipson, Breaking the delay-bandwidth limit in a photonic structure. *Nat. Phys.* **3**(6), 406-410 (2007), doi:10.1038/nphys600.
48. T. Kampfrath, D. M. Beggs, T. P. White, A. Melloni, T. F. Krauss, L. Kuipers, Ultrafast adiabatic manipulation of slow light in a photonic crystal. *Phys. Rev. A - At. Mol. Opt. Phys.* **81**(4), 043837 (2010), doi:10.1103/PhysRevA.81.043837.
49. D. M. Beggs, I. H. Rey, T. Kampfrath, N. Rotenberg, L. Kuipers, T. F. Krauss, Ultrafast tunable optical delay line based on indirect photonic transitions. *Phys. Rev. Lett.* **108**(21), 213901 (2012), doi:10.1103/PhysRevLett.108.213901.
50. M. Castellanos Muñoz, A. Y. Petrov, L. O'Faolain, J. Li, T. F. Krauss, M. Eich, Optically induced indirect photonic transitions in a slow light photonic crystal waveguide. *Phys. Rev. Lett.* **112**(5), 053904 (2014), doi:10.1103/PhysRevLett.112.053904.
51. F. Monticone, H. M. Doeleman, W. Den Hollander, A. F. Koenderink, A. Alù, Trapping Light in Plain Sight: Embedded Photonic Eigenstates in Zero-Index Metamaterials. *Laser Photonics Rev.* **12**(5), 1700220 (2018), doi:10.1002/lpor.201700220.
52. S. V. Silva, T. A. Morgado, M. G. Silveirinha, Multiple embedded eigenstates in nonlocal plasmonic nanostructures. *Phys. Rev. B* **101**(4), 041106 (2020), doi:10.1103/PhysRevB.101.041106.
53. O. Reshef, I. De Leon, M. Z. Alam, R. W. Boyd, Nonlinear optical effects in epsilon-near-zero media. *Nat. Rev. Mater.* **4**(8), 535-551 (2019), doi:10.1038/s41578-019-0120-5.
54. J. B. Khurgin, M. Clerici, V. Bruno, L. Caspani, C. DeVault, J. Kim, A. Shaltout, A. Boltasseva, V. M. Shalaev, M. Ferrera, D. Faccio, N. Kinsey, Adiabatic frequency shifting in epsilon-near-zero materials: the role of group velocity. *Optica* **7**(3), 226-231 (2020), doi:10.1364/optica.374788.
55. N. Kinsey, C. DeVault, J. Kim, M. Ferrera, V. M. Shalaev, A. Boltasseva, Epsilon-near-zero Al-doped ZnO for ultrafast switching at telecom wavelengths. *Optica* **2**(7), 616-622 (2015), doi:10.1364/optica.2.000616.
56. M. Z. Alam, I. De Leon, R. W. Boyd, Large optical nonlinearity of indium tin oxide in its epsilon-near-zero region. *Science* **352**(6287), 795-797 (2016), doi:10.1126/science.aae0330.
57. Y. Zhou, M. Z. Alam, M. Karimi, J. Upham, O. Reshef, C. Liu, A. E. Willner, R. W. Boyd, Broadband frequency translation through time refraction in an epsilon-near-zero material. *Nat. Commun.* **11**(1), 1-7 (2020), doi:10.1038/s41467-020-15682-2.
58. C. F. Bohren, D. R. Huffman, *Absorption and Scattering of Light by Small Particles* (Wiley, 1998).
59. Lumerical FDTD Solutions, http://www.lumerical.com/.
60. C. A. Balanis, *Advanced Engineering Electromagnetics* (Wiley, 2012).
61. A. M. Shaltout, V. M. Shalaev, M. L. Brongersma, Spatiotemporal light control with active metasurfaces. *Science* **364**(6441), eaat3100 (2019), doi:10.1126/science.aat3100.
62. B. Kang, J. H. Woo, E. Choi, H.-H. Lee, E. S. Kim, J. Kim, T.-J. Hwang, Y.-S. Park, D. H. Kim, J. W. Wu, Optical switching of near infrared light transmission in metamaterial-liquid crystal cell structure. *Opt. Express* **18**(16), 16492-16498 (2010), doi:10.1364/oe.18.016492.
63. S. Xiao, U. K. Chettiar, A. V. Kildishev, V. Drachev, I. C. Khoo, V. M. Shalaev, Tunable magnetic response of metamaterials. *Appl. Phys. Lett.* **95**(3), 033115 (2009), doi:10.1063/1.3182857.

64. M. Decker, C. Kremers, A. Minovich, I. Staude, A. E. Miroshnichenko, D. Chigrin, D. N. Neshev, C. Jagadish, Y. S. Kivshar, Electro-optical switching by liquid-crystal controlled metasurfaces. *Opt. Express* **21**(7), 8879-8885 (2013), doi:10.1364/oe.21.008879.

(†) Online link to supplementary movie 1: http://tiny.cc/supp_mov1 and supplementary movie 2: http://tiny.cc/supp_mov2

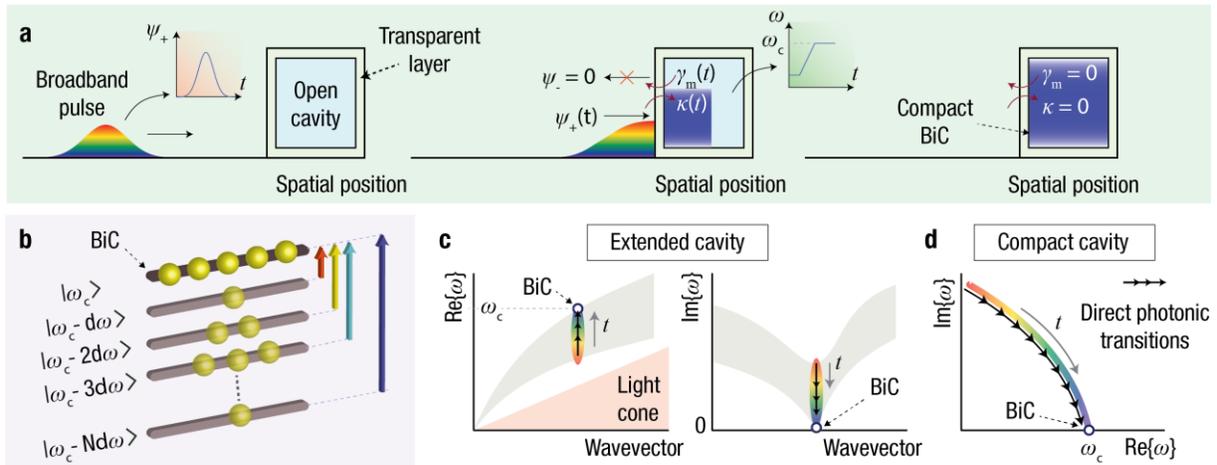

**Figure 1 | Broadband light trapping into a radiationless open cavity eigenstate. a,** Illustration of the complete capturing of a broadband incident pulse into a time-varying open cavity supporting a bound state in the continuum (BiC). The temporal modulation is designed such that light is coupled into the lossless resonator with no reflections (critical coupling condition). **b,** Temporally modulated systems do not conserve energy locally and, therefore, can lead to frequency transitions between optical states (the number of photons, represented by yellow spheres, is conserved). This effect can be used to 'force' a broadband pulse into a BiC. **c,** Illustrative complex dispersion diagram of an extended open cavity; **d,** complex frequency diagram of a bounded resonator. The rainbow-colored regions together with the arrows conceptually represent the trajectory along which the frequency transitions occur in order to capture a light pulse with non-zero bandwidth into a BiC (shown with white circular markers). We stress that, at the BiC condition, the structure is effectively closed only from the point of view of the specific mode of interest, while remaining open for generic excitation fields.

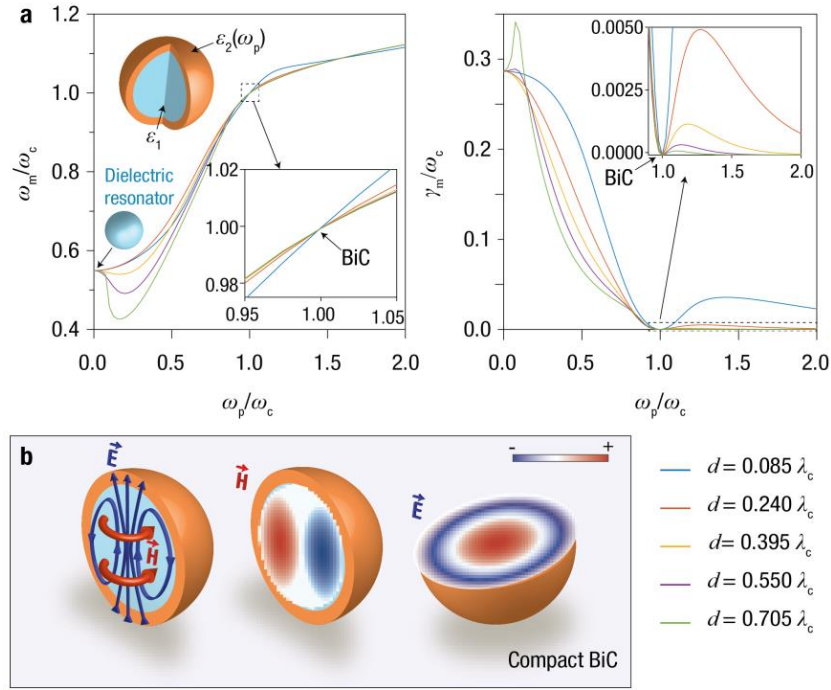

**Figure 2 | Tunability of the complex resonance eigenfrequency. a,** Complex resonance eigenfrequency ($\omega_0 = \omega_m + i\gamma_m$) of the relevant mode of a core-shell scatterer as a function of the plasma frequency of the shell, and for different values of the shell thickness. The lower right insets show an enlarged view around the BiC frequency ($\omega_c$). **b,** Illustrative magnetic and electric field vectors (left), magnetic field profile (middle) and electric field profile (right) of the radiationless eigenmode of the core-shell scatterer (corresponding to a BiC at $\omega_c$). The middle and right panels show time-snapshots of the field component orthogonal to the cross section. The dielectric core has a relative permittivity of 5 and radius equal to $R = 0.320\,\lambda_c$, where $\lambda_c$ is the free space wavelength at the BiC frequency; the plasmonic shell has thickness $d$ and follows a standard Drude dispersion model with vanishing absorption.

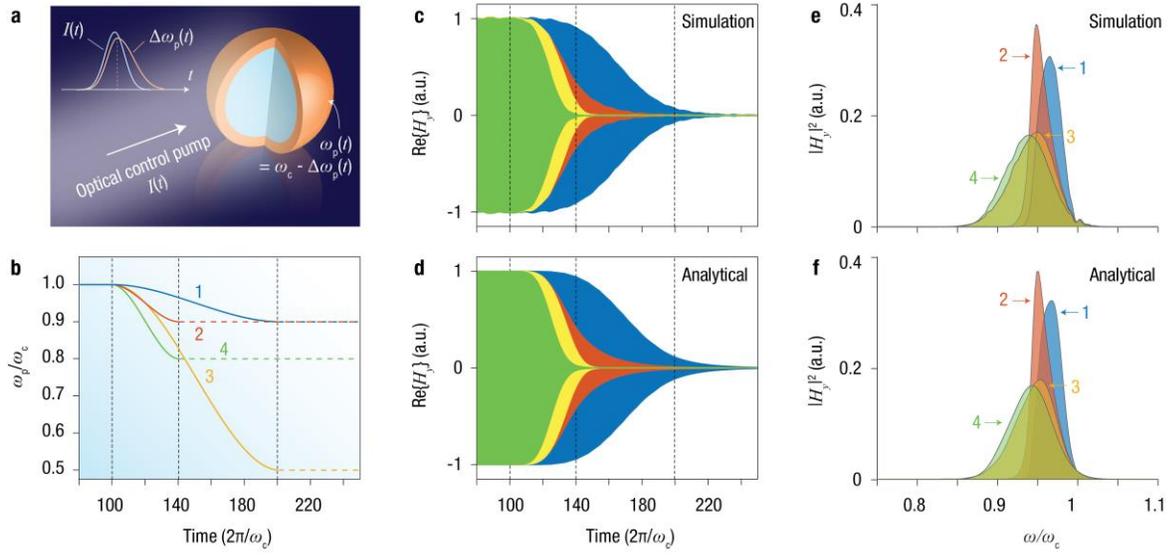

**Figure 3 | Wave dynamics of an initially excited open cavity subject to a temporal modulation. a,** Conceptual representation of a core-shell scatterer dynamically tuned through a time-varying optical pump. The upper left inset illustratively shows the change of the shell plasma frequency $\Delta\omega_p$, which can be induced, for instance, through interband or intraband absorption processes in the presence of a strong optical control pump $I(t)$ [53-57]. **b,** Different examples of temporal modulations of the shell plasma frequency. **c,d,** Numerically obtained (c) (through FDTD simulations) and analytically calculated (d) time-domain response of the magnetic field (y-component, $H_y$) inside an initially excited open resonator subject to the dynamic modulations in panel (b). The scatterer is initially tuned at the BiC condition, i.e., $\omega_p=\omega_c$. The radiationless dipolar mode of interest is excited by an ideal point source inside the structure before the modulation starts. **e,f,** Numerically (e) and analytically (f) calculated spectra of the radiated magnetic field intensity, measured at $(x, y, z) = (0.65, 0, 0)\ \lambda_c$ (relative to the center of the resonator). The structural parameters of the core-shell resonator are the same as in Fig. 2 with $d = 0.085\ \lambda_c$.

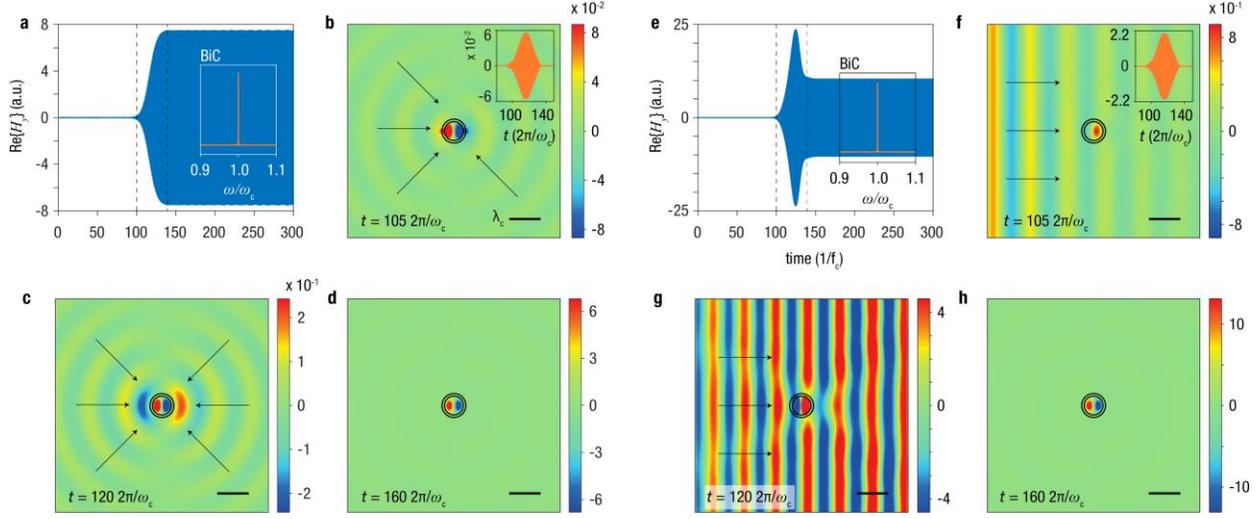

**Figure 4 | Broadband excitation of a BiC in a time-varying open cavity. a-h,** External excitation of a BiC by a spherical-wave pulse (TM$^r$ spherical harmonic of order 1) (a-d) and a plane-wave pulse (e-h). (a,e) Time-domain response of the magnetic field inside the resonator at $(x, y, z) = (0.14, 0, 0)\ \lambda_c$ (relative to the center of the resonator). The inset shows the BiC field spectrum. (b-d,f-h) Time-snapshots of the $H_y$ field distribution on the center $xy$-plane at time steps $t = 105\ 2\pi/\omega_c$ (b,f), $t = 115\ 2\pi/\omega_c$ (c,g) and $t = 160\ 2\pi/\omega_c$ (d,h). The inset of panels (b),(f) shows the magnetic field temporal profile of the impinging wave at $(x, y, z) = (-3.4, 0, 0)\ \lambda_c$. The structural parameters of the considered scatterer are the same as in Fig. 3, and the applied temporal modulation is the same as the modulation numbered "4" in Fig. 3, but time reversed around time step $t = 120\ 2\pi/\omega_c$. Scale bars correspond to $\lambda_c$, black lines denote the structural boundaries of the core-shell scatterer, and black arrows indicate the direction of wave propagation. Time-domain animations of the spherical- and plane-wave excitation cases are available as supplementary movies 1 and 2$^\dagger$, respectively, which provide a detailed look into the response of the structure under broadband illumination.

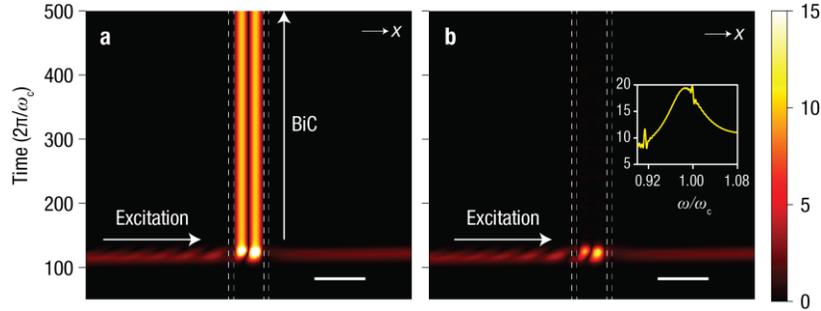

**Figure 5 | Broadband BiC excitation with and without temporal modulation. a,b,** Time-resolved magnetic field amplitude distribution, along the center $x$-axis, for plane-wave excitation of the considered scatterer with (a) and without (b) dynamic temporal modulation. Only in the time-varying case the BiC is excited by the broadband incident wave. In the static case, the internal fields are non-zero, but do not induce long-lived resonances. The inset of panel (b) shows the spectrum of the internal fields. The dynamic modulation in (a) is the same as in Fig. 4, while in (b) the shell plasma frequency is static and equal to $\omega_c$. The temporal profile of the plane wave pulse is the same as in the inset of Fig. 4(b). Scale bars correspond to $\lambda_c$ and white dashed lines indicate the structural boundaries.

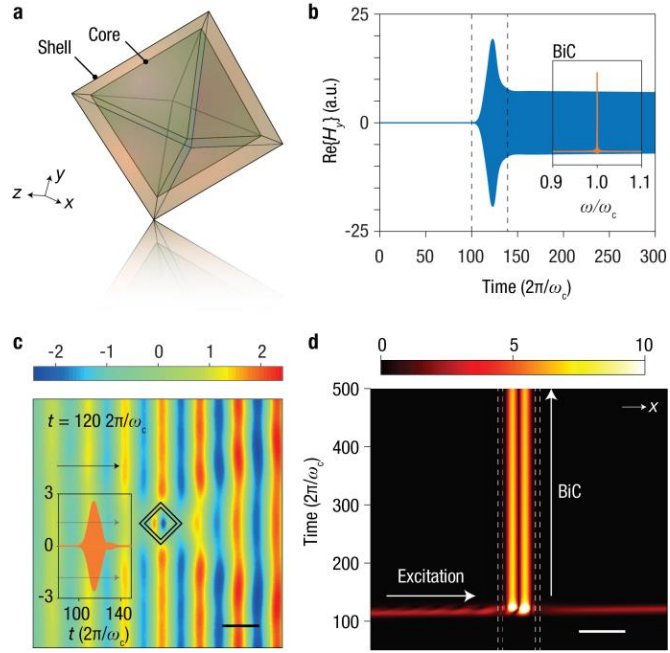

**Figure 6 | Broadband excitation of a BiC in a time-varying open cavity with noncanonical geometry. a,** Schematic of a core-shell resonant scatterer made of two concentric regular octahedrons. **b,** Temporal variation of the magnetic field, $H_y$, inside the resonator under broadband plane-wave excitation. The inset shows the BiC field spectrum. **c**, Time-snapshot of the $H_y$ field distribution on the center $xy$- plane at $t = 120\ 2\pi/\omega_c$. Inset: $H_y$ temporal profile of the impinging plane wave at the same position as in Fig. 4. **d**, Time-resolved magnetic field amplitude distribution along the center $x$-axis, similar to Fig. 5(a). The edge lengths of the inner and outer regular octahedrons are $0.68\ \lambda_c$ and $0.88\ \lambda_c$, respectively. The material properties and the applied dynamic modulation are the same as in Figs. 4 and 5. Scale bars correspond to $\lambda_c$; black lines in (c) and white dashed lines in (d) denote the structural boundaries of the resonator.